\pdfoutput=1
\documentclass[amsmath,prd,aps,twocolumn,showpacs,superscriptaddress,floatfix,nofootinbib]{revtex4}
\usepackage{float}
\usepackage{amsfonts}
\usepackage{bm}
\usepackage{subfigure}

\usepackage{color}
\usepackage{graphicx}
\definecolor{darkgreen}{cmyk}{0.85,0.2,1.00,0.2}
\newcommand{\be}{\begin{equation}}
\newcommand{\ee}{\end{equation}}
\newcommand{\ba}{\begin{eqnarray}}
\newcommand{\ea}{\end{eqnarray}}

\newcommand\lsim{\mathrel{\rlap{\lower4pt\hbox{\hskip1pt$\sim$}}
        \raise1pt\hbox{$<$}}}
\newcommand\gsim{\mathrel{\rlap{\lower4pt\hbox{\hskip1pt$\sim$}}
        \raise1pt\hbox{$>$}}}

\def\d{{\bf d}}
\def\t{{\bf t}}

\def\n{{\bf \widehat n}}

\begin{document}

\title{A WISE measurement of the ISW effect}

\author{Simone Ferraro}
\affiliation{Department of Astrophysical Sciences, Peyton Hall, Princeton University, Princeton, NJ 08544 USA }
\author{Blake D. Sherwin}
\affiliation{Department of Physics, University of California, Berkeley, CA, USA}
\affiliation{Miller Institute for Basic Research in Science, University of California, Berkeley, CA, USA}
\author{David N. Spergel}
\affiliation{Department of Astrophysical Sciences, Peyton Hall, Princeton University, Princeton, NJ 08544 USA}

\date{\today}


\begin{abstract}
The \textit{Integrated Sachs-Wolfe effect} (ISW) measures the decay of the gravitational potential due to cosmic acceleration and is thus a direct probe of Dark Energy. In some of the earlier studies, the amplitude of the ISW effect was found to be in tension with the predictions of the standard $\Lambda$CDM model. We measure the cross-power of galaxies and AGN from the WISE mission with CMB temperature data from WMAP9 in order to provide an independent measurement of the ISW amplitude.  Cross-correlations with the recently released Planck lensing potential maps are used to calibrate the bias and contamination fraction of the sources, thus avoiding systematic effects that could be present when using auto-spectra to measure bias. We find an amplitude of the cross-power of $\mathcal{A}  = 1.24\pm 0.47$ from the galaxies and $\mathcal{A} = 0.88 \pm 0.74$ from the AGN, fully consistent with the $\Lambda$CDM prediction of $\mathcal{A} =1$.  The ISW measurement signal-to-noise ratio is 2.7 and 1.2 respectively, giving a combined significance close to $3 \sigma$. Comparing the amplitudes of the galaxy and AGN cross-correlations, which arise from different redshifts, we find no evidence for redshift evolution in Dark Energy properties, consistent with a Cosmological Constant.\end{abstract}

\maketitle


\section{Introduction}
\label{sec:intro}

The nature and the properties of Dark Energy are among the most significant unsolved problems in physics.  We now know that Dark Energy accounts for about 70\% of the energy density of the Universe and is causing the cosmic expansion to accelerate. Measurements of Type Ia supernovae, Baryon Acoustic Oscillations, galaxy clusters or gravitational lensing (of the Cosmic Microwave Background, galaxies, or strongly lensed quasars) \cite{Hicken, Kessler, Reid, Riess, Vikhlinin, Suyu}, when combined with measurements of the Cosmic Microwave Background (CMB) anisotropies \cite{WMAP9, Planck1}, all provide evidence for an accelerated expansion \cite{SN} and imply a flat and Dark Energy dominated universe.

While theorists have proposed a large number of models to explain cosmic acceleration, including modifications to General Relativity on large scales \cite{MG}, there exist only very few observational windows into the properties of this phenomenon.

Measurements of the \textit{Integrated Sachs-Wolfe} effect (ISW) \cite{SW} provide a powerful method to probe Dark Energy, as this effect is sensitive to the time evolution of the gravitational potential sourced by Large Scale Structure and thus probes Einstein's equations beyond the simple Friedmann equation. The ISW effect is the distortion of the CMB temperature due to the time evolution of the gravitational potential along the line of sight: photons entering a gravitational potential well blue-shift and subsequently redshift when leaving the well. In a matter-dominated universe, the gravitational potential is time-independent on large scales, so the amount of blue- and red- shifting is the same and the photon energy is overall unchanged; Dark Energy causes an accelerated expansion, making the gravitational potential shallower with time and resulting in a net blue-shift of the photons. This effect is too small to be detected directly in the CMB spectrum \cite{Huffen} but it is expected to be measurable through the correlation between the measured temperature anisotropies of the CMB and the Large Scale Structure, which acts as a tracer of the gravitational potential.  

Such analyses have been carried out in earlier work (see for example \cite{Scranton, Fosalba, Granett, Pad, Gian08, Gian12, Gian13, BOSS, Goto, Goto2}), with the strongest detection to date (at the 4.5$\sigma$ level) relying on the combination of many different data sets \cite{Gian08}. One interesting feature of several of the previous studies is that the cross-correlation signal lies systematically above (by $\sim 1 - 2 \sigma$) the predicted value in the standard cosmological model (in which the Dark Energy is a Cosmological Constant) \cite{Gian08, Goto, Ho}. The same is true for some analyses based on stacking large clusters and voids \cite{Granett, PlanckISW}.

We perform a new ISW cross-correlation analysis using a sample of galaxies and quasars from the \textit{Wide-field Infrared Survey Explorer} (WISE, \cite{WISE}), which scanned the full sky in 4 frequency bands, ranging from 3.4 to 22 $\mu$m, and detected hundreds of millions of sources. The 3.4 $\mu$m band probes massive galaxies out to $z \sim 1$ and with a median redshift of $0.3$ \cite{Yan}. The large area of the survey, together with its redshift distribution and the large number of sources, makes WISE one of the best catalogs for this kind of work. An early study with the WISE preliminary release catalog found an amplitude that is 2$\sigma$ above the $\Lambda$CDM prediction \cite{Goto}, while a subsequent work \cite{Goto2} using the full sky galaxy catalog found an amplitude consistent with $\Lambda$CDM, but at low significance ($1\sigma$). We use a larger sample (applying less restrictive cuts to the data) with higher median redshift and expect to detect the signal at a considerably higher significance.

Since galaxies and quasars trace the dark matter and hence the potential up to a bias factor (a proportionality factor relating tracer overdensity to mass overdensity), it is crucial to measure the bias reliably in order to be able to compare the ISW amplitude with theoretical predictions. Methods to measure the bias from the auto-correlation spectrum can be prone to systematic errors, especially for WISE maps which contain strong galactic and instrumental signals, and can lead to incorrect conclusions about the amplitude of the ISW effect. Recent progress in the measurement of the gravitational lensing of the CMB by the \textit{Atacama Cosmology Telescope} (ACT, \cite{ACT, BlakeLens}), \textit{South Pole Telescope} (SPT, \cite{SPT, SPTLens, SPTpol}), POLARBEAR \cite{PB1, PB2} and the Planck Satellite \cite{Planck1, PlanckLens}, allow a direct measurement of bias (lensing is sourced directly by the gravitational potential itself), by cross-correlating lensing potential maps with the tracer field. We expect this measurement to be more robust and less prone to systematic errors.

The paper is organized as follows: The ISW effect is briefly reviewed in section \ref{sec:ISW}. Sections \ref{sec:WISE} and \ref{sec:CMB} introduce our tracer and CMB datasets, while in section \ref{sec:lens} we discuss the calibration of the bias using CMB lensing. Our ISW results are presented in section \ref{sec:res}, followed by a discussion and conclusions in section \ref{sec:conclusions}.
\vspace{1cm}

\section{The ISW effect}
\label{sec:ISW}

As discussed in the introduction, the ISW effect is a secondary CMB anisotropy which is due to the time variation of the gravitational potentials along the line of sight \cite{SW}:
\begin{align}
\left( \frac{\Delta T}{T} \right)_{ISW}(\n) &= - \int d \eta \ e^{-\tau(z)} (\dot{\Phi} + \dot{\Psi}) [\eta, \n (\eta_0 - \eta)] \nonumber \\
\ & \approx - 2 \int d \eta \ \dot{\Phi} [\eta, \n (\eta_0 - \eta)]
\end{align}
where in the second line we have used the GR prediction that in absence of anisotropic stresses $\Phi = \Psi$ and have approximated the optical depth $\tau(z) \ll 1$ over the period where $\dot{\Phi} \neq 0$, so that we can take $e^{-\tau(z)} \approx 1$.
Note that during matter domination, $\dot{\Phi} = 0$ and there in no ISW contribution. Since in the standard cosmological model the effect of Dark Energy is relevant only at $z \lsim 1$,  the largest contribution comes from the very largest scales. The typical low $\ell$ contribution to the CMB fluctuation spectrum is $\sim 100\ \mu$K$^2$, compared to the $\sim 1000\ \mu$K$^2$ of the primary fluctuations, too small to be detected directly in presence of cosmic variance.
This problem can be overcome by cross-correlating the observed CMB temperature with tracers of the gravitational potential, such as galaxies or quasars, that would otherwise be uncorrelated with the CMB in the absence of the ISW contribution.

We will work with the projected overdensity field of tracers (galaxies or quasars), which can be expressed in terms of the matter overdensity $\delta$:
\be
\delta_g(\n) = \int dz \ b(z) \frac{dN}{dz} \delta(\n, z)
\ee
Where we have assumed a (redshift dependent) linear bias model for the tracers and $dN/dz$ is the redshift distribution normalized such that $\int dz' \frac{dN}{dz'} = 1$.

We can compute the angular cross-correlation\footnote{Here we assume that the ISW contribution is the only component of the temperature anisotropy correlated with low-redshift tracers of the potential, so that we can write $C_{\ell}^{T g} = C_{\ell}^{\dot{\Phi}g}$.}:
\be
C_{\ell}^{T g} = C_{\ell}^{\dot{\Phi}g} = 4 \pi \int \frac{dk}{k} \Delta^2_m(k) K_{\ell}^{\dot{\Phi}}(k) K_{\ell}^g (k)
\ee
in terms of the dimensionless (linear) matter power spectrum at redshift $z= 0$, $\Delta^2_m(k) = k^3 P(k, z=0) /2 \pi^2$.
Here the galaxy and ISW weight functions are given by:
\begin{align}
K_{\ell}^{g}(k) & =  \int dz \ b(z) \frac{dN}{dz} D(z) \ j_{\ell}[k \chi(z)] \\
K_{\ell}^{\dot{\Phi}}(k) & =  \frac{3 \Omega_m H_0^2}{k^2} \int dz \frac{d}{dz} ( (1+z) D(z) )\  j_{\ell}[k \chi(z)]
\end{align}
where $j_{\ell}$ are the spherical Bessel functions, $D(z)$ is the linear growth factor normalized to $D(z=0) = 1$ and $\chi(z) = \eta_0 - \eta(z)$ is the comoving distance to redshift $z$.

A simple Fisher matrix analysis gives the expected signal-to-noise ratio for a coverage fraction $f_{sky}$,
\begin{align}
\left( \frac{S}{N} \right)^2 & \approx f_{sky} \sum_{\ell} (2 \ell +1) \frac{[C_{\ell}^{Tg}]^2}{C_{\ell}^{TT} C_{\ell}^{gg} + [C_{\ell}^{Tg}]^2} \\
& \approx  f_{sky} \sum_{\ell} (2 \ell +1) \frac{[C_{\ell}^{Tg}]^2}{C_{\ell}^{TT} C_{\ell}^{gg} }
\end{align}
where in the second line we have used the fact that the correlation is weak $C_{\ell}^{Tg} \ll \sqrt{C_{\ell}^{TT} C_{\ell}^{gg}}$. 
It can be shown that most of the signal-to-noise comes from $\ell \sim 20$ and $z \sim 0.4$, with a wide redshift distribution, and that the contributions from $z > 1.5$ and $\ell > 100$ are negligible \cite{Afshordi}.

Note that due to cosmic variance, there is a theoretical maximum for the signal-to-noise ratio, which we can see as follows: The correlation coefficient $r \equiv C_{\ell}^{\dot{\Phi}g} / \sqrt{C_{\ell}^{\dot{\Phi}\dot{\Phi}} C_{\ell}^{gg}}$ is constrained to be $-1 \leq r \leq 1$, so that
$[C_{\ell}^{Tg}]^2 = [C_{\ell}^{\dot{\Phi}g}]^2  \leq C_{\ell}^{\dot{\Phi}\dot{\Phi}} C_{\ell}^{gg}$.
Therefore
\begin{align}
\left( \frac{S}{N} \right)^2 & \approx f_{sky} \sum_{\ell} (2 \ell +1) \frac{[C_{\ell}^{Tg}]^2}{C_{\ell}^{TT} C_{\ell}^{gg}} \\
& \leq f_{sky} \sum_{\ell} (2 \ell +1) \frac{C_{\ell}^{\dot{\Phi}\dot{\Phi}}}{C_{\ell}^{TT}}
\end{align}
This can be evaluated in a given cosmological model and for $\Lambda$CDM we find $(S/N) \lsim 7.6 \sqrt{f_{sky}}$
and about 15\% more if we add polarization information \cite{Gian12}.
\section{WISE data}
\label{sec:WISE}
WISE scanned the entire sky in four bands at 3.4, 4.6, 12 and 22 $\mu$m (W1 to W4) and provided a much deeper dataset than other experiments at similar frequencies (such as 2MASS and IRAS). The WISE W1 and W2 bands primarily probe starlight coming from other galaxies or galactic stars, while the W3 and W4 bands are more sensitive to the thermal emission from dust grains.  

The WISE Source Catalog \cite{WISE} contains more than 500 million sources which are detected at $S/N >$ 5 in at least one band (usually W1 since it is the most sensitive). Galactic stars and quasars each account for approximately 12\% of the catalog at high galactic latitude. Approximately $70\%$ are normal star-forming galaxies, while $6\%$ are unusually red, unidentified sources  \cite{Yan}. Previous work \cite{Yan, Jarrett, Stern12} has shown that the four WISE bands are sufficient to effectively distinguish stars and quasars from normal galaxies. The details of this color-color selection are outlined in the next subsections.

Unfortunately, parts of the WISE catalog are contaminated by moonlight: when WISE observes near the Moon (or as far as 30 deg away), stray light can affect the images and produce spurious detections. This is visible as several bright (overdense) stripes, which are perpendicular to the ecliptic equator and parallel to the WISE scan direction. The catalog's {\tt{moon\_lev}} flag denotes the fraction of frames that are believed to be contaminated. We discard all objects that have {\tt{moon\_lev}} $ >4$ in any band and regions with high density of such objects are added to the mask. 

We also discard any source for which {\tt{cc\_flags}} $\neq 0$, since it is considered an artifact (diffraction spike, optical ghost, etc.). 

Due to the scan strategy, the coverage depth is very inhomogeneous (the poles were scanned to much greater depth than the equator) and the selection function is mostly unknown.
The median coverage in W1 is 15 exposures, with 12 exposure being the `typical' number for points near the equator and 160 for points near the ecliptic poles. Plotting the source magnitude distribution as a function of position of the sky, we find that for high galactic latitude, the distribution is fairly uniform for W1 $< 17.0$.  According to the WISE Explanatory Supplement\footnote{http://wise2.ipac.caltech.edu/docs/release/allsky/expsup/ \\ sec2\_2.html}, the catalog is 95\% complete for sources with W1 $< 16.6$. Therefore we apply this magnitude cut to ensure good completeness and uniformity and at the same time retain the largest number of sources.

Below we outline our selection criteria for stars, galaxies and AGN:
\subsection{Stars}
Emission from stars in the mid-IR is dominated by the Rayleigh-Jeans tail of the spectrum, meaning that the color is close to zero and approximately independent of surface temperature.
We use the following color cuts proposed in \cite{Jarrett} to separate stars from galaxies and AGN: W1 $< 10.5$, W2 $-$ W3 $ <1.5$ and W1 $-$ W2 $< 0.4$. In addition, we find that stars close to the galactic plane are effectively removed by classifying as `star' anything with W1 $-$ W2 $<0$.

Dust-poor elliptical galaxies at low redshift are hard to distinguish from stars with WISE colors alone and can therefore be misidentified and fall into this category.

\subsection{Galaxies}
Here we adopt an empirical definition of ``Galaxy'' as anything not classified as a star or AGN. Due to the negative k-correction in the IR, the WISE W1 band can probe galaxies out to $z \gsim 1$, since the W1 flux does not change significantly in the range $z \sim 0.5 - 1.5$ \cite{Yan}. We use the redshift distribution of WISE galaxies as measured in \cite{Yan}. In this paper, the authors cross-matched WISE sources with SDSS DR7 \cite{DR7} in high galactic latitude regions and found the distribution to be fairly broad, peaking at $z \sim 0.3$ and extending all the way to $z = 1$.  In order to more effectively remove galactic stars and be able to use a larger portion of the sky, we had to make the additional cut W1 $-$ W2 $>0$, compared to \cite{Yan}. The effect of this on the redshift distribution should be negligible, since from their color-color diagrams, the vast majority of galaxies are shown to indeed have W1 $-$ W2 $>0$.  To further test the effect of uncertainties in the redshift distribution, we repeat the analysis by shifting the whole distribution by $\Delta z = \pm 0.1$ (corresponding to a $\sim$30\% shift in the peak $z$) and find that the best fit ISW amplitude is only changed by $\sim 5\%$, corresponding to about 0.1$\sigma$. We therefore conclude that it is appropriate to use the distribution as in \cite{Yan} without additional corrections. 

The redshift distribution and the very large number of sources (our sample consists of approximately 50 million galaxies) make WISE nearly ideal for ISW cross-correlation.

The criterion W1 $-$ W2 $>0$ for galaxies ensures that the stellar contamination is small, at the cost of omitting a small number of galaxies. The remaining contamination, if uncorrelated with the CMB, affects the normalization of $C_{\ell}^{Tg}$ in the same way as it affects the cross-correlation with CMB lensing, and therefore can be calibrated out (see section \ref{sec:lens}). If in addition the contamination sources are clustered (like stars close to the galactic plane), they will add to the auto-power spectrum on large scales, thus lowering the statistical significance of the ISW measurement.
 
A bias model that is constant with redshift is expected to be appropriate for WISE selected galaxies; we measure the bias via lensing in section \ref{sec:lens}. To investigate the dependence of our results on the uncertainties in bias evolution, we also repeat the analysis for an evolving bias model $b^G(z) = b_0^{G}(1+z)$, with constant $b_0^G$.

Our conservative masking leaves $f_{sky} = 0.47$ and about 50 million galaxies.

\begin{figure}[ht]
\includegraphics[width=8.0cm]{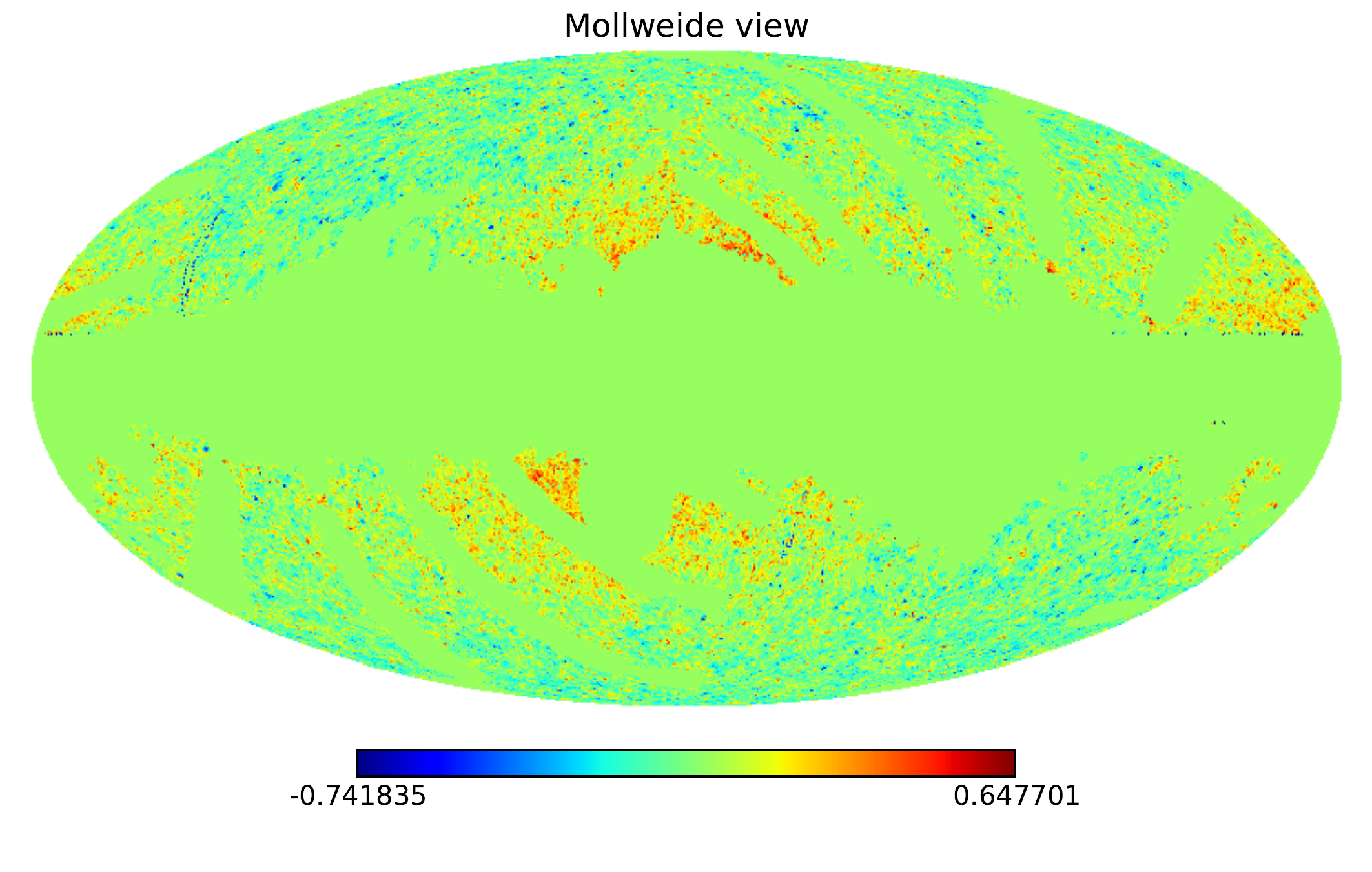}
 \caption{The WISE galaxy overdensity map, including the mask, where the overdensity is 0.}\label{galsplot}
\end{figure}

\subsection{AGN}
The mid-IR selection of AGN is a well-studied problem.  Following \cite{Assef12, Stern12} we use the selection criteria W1 $-$ W2 $> 0.85$ and W2 $<15.0$. This has been shown to work well for both Type 1 and Type 2 AGN up to redshift $z \lsim 3$ and leads to a source density of 42 deg$^{-2}$. Mid-IR selection is not significantly affected by dust extinction and the only potential contaminants are brown dwarfs and asymptotic giant branch stars, both of which have much smaller surface density. 

We use the redshift distribution of WISE AGN that has been recently measured in \cite{Geach13} by cross-matching AGN on 7.9 deg$^2$ of the Bo{\"o}tes/AGES field. The authors show that it peaks at $z \sim 1.1$, with a spread $\Delta z \sim 0.6$ and further constrain the contamination fraction to be less than 15\%.

We take the redshift dependence of the bias to be the one appropriate for the Type 1 QSOs, as suggested by \cite{Croom}: $b^A(z) = b^A_0 [0.53 + 0.289(1+z)^2]$, where $b_0$ is an overall amplitude, which we measure from the cross-correlation with CMB lensing maps.

Stellar contamination is expected to be very small, since AGN are easily distinguishable from stars using WISE bands.

Our masking leaves $f_{sky} = 0.48$ and about 910,000 AGN.

\begin{figure}[ht]
\includegraphics[width=8.0cm]{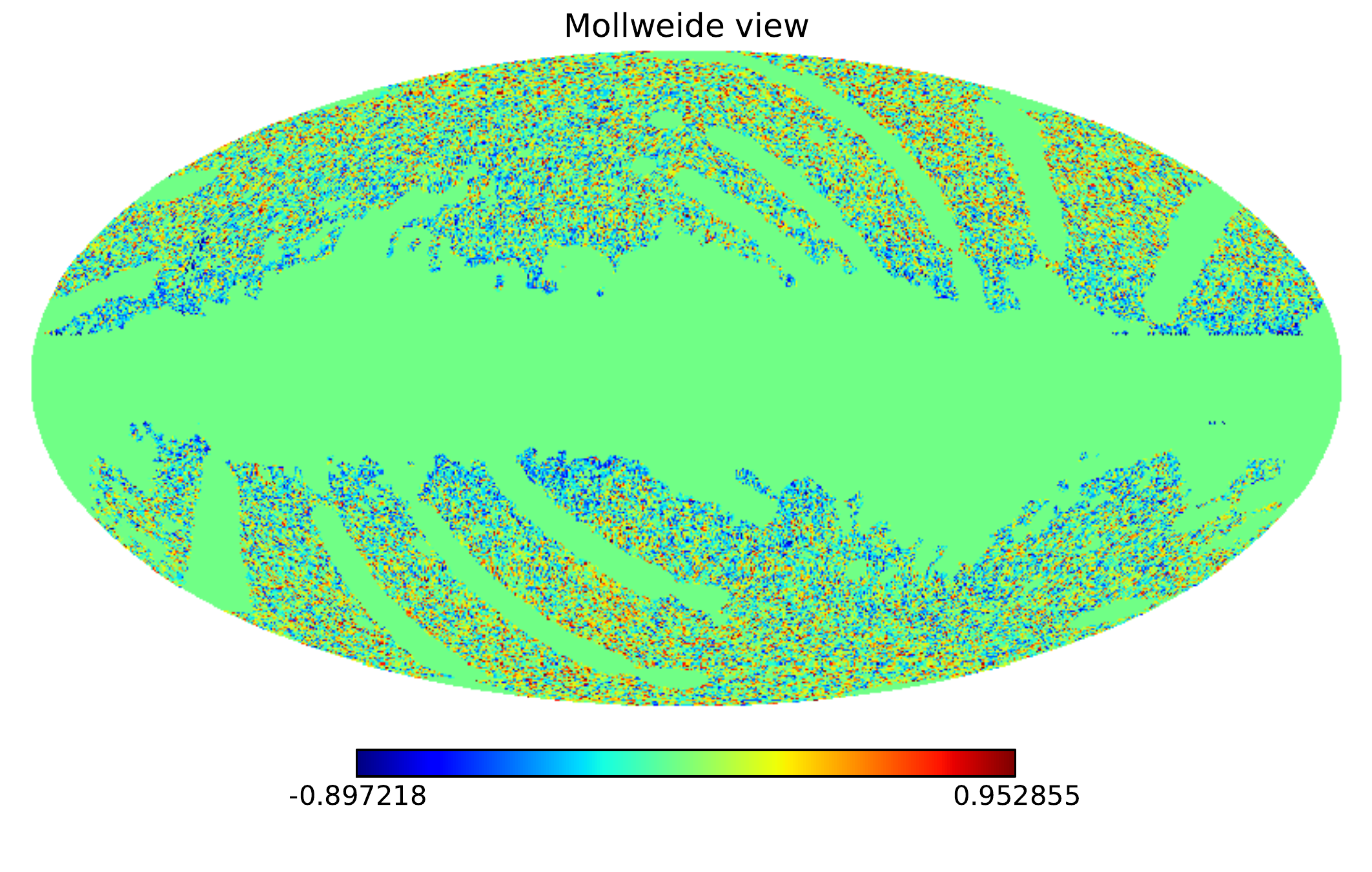}
 \caption{The WISE AGN overdensity map, including the mask, where the overdensity is 0.}\label{qsoplot}
\end{figure}

\section{CMB data}
Our CMB temperature data is obtained from the foreground reduced WMAP9 maps \cite{WMAP9} in the Q, V and W bands (respectively at 40, 60 and 90 GHz). At the scales of interest ($\ell \lsim 100$), the data is cosmic variance limited with negligible instrumental noise. For the CMB we apply the {\tt KQ75y9} extended temperature analysis mask, which includes point sources detected in WMAP and has $f_{sky} \approx 0.69$. The total mask is the product of the CMB mask and the appropriate WISE mask for AGN or galaxies. The same comprehensive mask is applied to both datasets before the cross-correlation analysis.
\label{sec:CMB}

\section{Lensing bias calibration}
We use weak lensing of the CMB by our tracers to measure an effective bias, which takes into account the level of contamination by stars or artifacts.
\label{sec:lens}

\subsection{Introduction}
The observed (lensed) temperature $T(\n)$ in a given direction $\n$ is a remapping of the original temperature $T_{\rm or}$ in the direction $\n + \d$, where $\d$ is the displacement field: $T(\n) = T_{\rm or}(\n + \d)$. 

It is convenient to work with the convergence field, defined as $\kappa \equiv - \nabla \cdot \d /2$ and which can be expressed as an integral along the line of sight  \cite{CoorayHu, Lewis}:
\be
\label{eq:conv}
\kappa(\n) = \int dz \ K^{\kappa}(z) \delta(\n, z)
\ee
In a flat universe (an assumption that we make throughout), the lensing kernel is given by:
\be
K^{\kappa}(z) = \frac{3 \Omega_m H_0^2}{2H(z)}  (1+z) \chi(z)  \frac{\chi_{*} - \chi(z)}{\chi_{*}}
\ee
where $\chi_{*} \sim 14$ Gpc is the comoving distance to the last scattering surface.

The cross-correlation between the lensing convergence and the projected density field can be calculated using the Limber approximation, which is expected to work well here, since we only use modes $\ell \gsim 50$:
\be
C_{\ell}^{\kappa g} = \int dz \ \frac{H(z)}{\chi^2(z)} \ K^{g}(z) K^{\kappa}(z) P\left(k = \frac{\ell + 1/2}{\chi(z)}, z \right)
\ee

We note that the linear bias factor $b(z)$ appears in $C_{\ell}^{\kappa g}$ and $C_{\ell}^{T g}$ weighed by different kernels and therefore it is important to account for the redshift dependence of $b$. As discussed previously, our fiducial galaxy bias is a constant with redshift, but we also investigate the model $b^G(z) = b_0^{G}(1+z)$, while for the AGN we take $b^A(z) = b^A_0 [0.53 + 0.289(1+z)^2]$.

\subsection{Planck lensing potential}
The Planck collaboration released a map of the lensing potential $\phi_L$ (related to the lensing convergence by $\kappa = -\nabla^2 \phi_L/2$), covering over 70\% of the sky. As we can see from equation (\ref{eq:conv}), this is a direct measurement of the projected density field out to the surface of last scattering, weighted by a broad kernel which peaks at $z \sim 2$.

The correlation between WISE and the Planck lensing potential was recently investigated in \cite{PlanckLens, Geach13}, where a $\sim 7 \sigma$ detection was found for both galaxies and quasars. Here we repeat the analysis with the same maps and masks used for the ISW work.

\subsection{Results}
\label{sec:res}

We use the Planck lensing potential and WISE maps at HEALPix \cite{Healpix} resolution $N_{side} = 512$ and measure the cross-correlation signal for $100 \leq \ell \leq 400$, correcting for the effects of the pixel window function and of the mask. Note that we use the same $\ell_{max}$ as in the cosmological analysis by the Planck team \cite{PlanckLens}. Including higher $\ell$ would probe the non-linear regime, where a constant bias model is likely to be inadequate and require corrections.  Furthermore, including higher $\ell$ would be unnecessary from a statistical point of view, as the error on the bias is not the dominant source of uncertainty on the ISW amplitude. Lacking realistic simulated Planck lensing maps, the error bars are computed from the variance of the values in a given $\ell$ bin. We have however checked that they are consistent but $\sim 30-60\%$ larger than the theory error bars computed in the Gaussian approximation, which represent a theoretical lower bound.

\begin{figure}[ht]
\includegraphics[width=8.5cm]{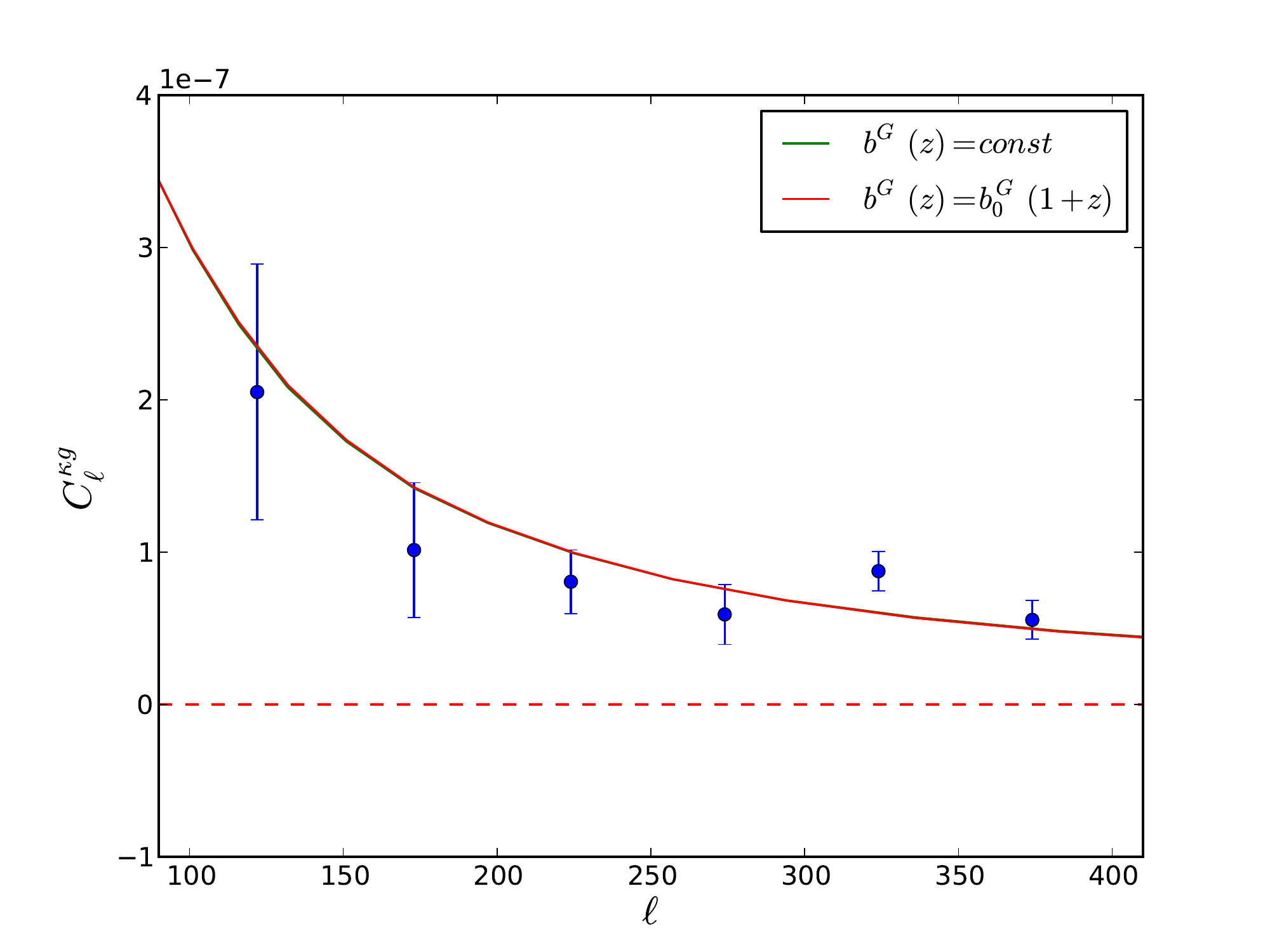}
 \caption{Lensing convergence-galaxy cross-correlation as a measure of the linear bias for WISE galaxies.}\label{galslens}
\end{figure}

\begin{figure}[ht]
\includegraphics[width=8.5cm]{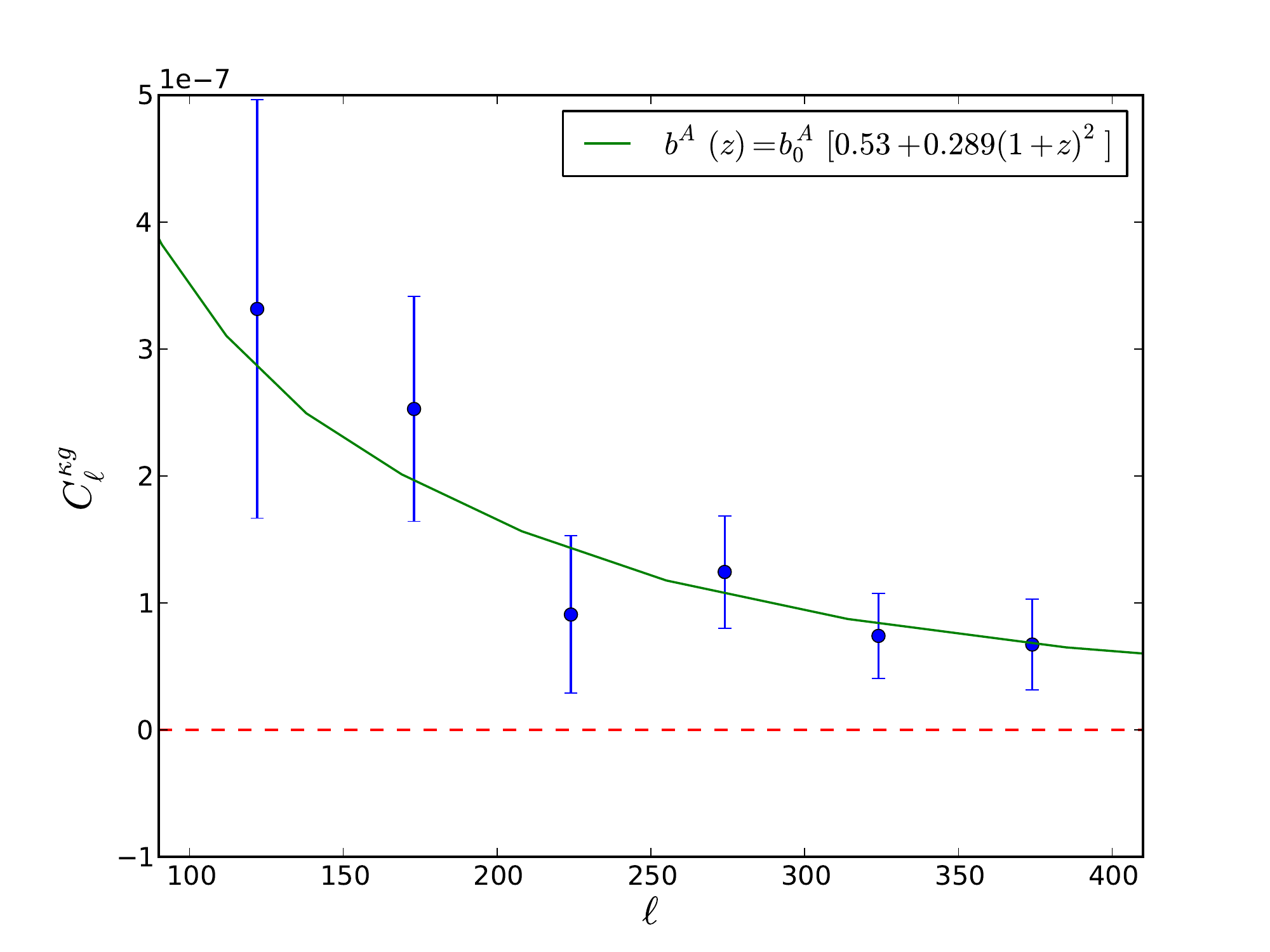}
 \caption{Lensing convergence-AGN cross-correlation as a measure of the linear bias for WISE AGN.}\label{qsolens}
\end{figure}

Figures \ref{galslens} and \ref{qsolens} show the cross-correlation signal. For our fiducial galaxy bias model (a redshift independent constant), we find $b^G = 1.41 \pm 0.15$. This value is larger than that found by the Planck Collaboration \cite{PlanckLens}; this difference is expected, because Planck uses a more conservative magnitude cut and hence measures bias of lower redshift, less biased sources. If instead we consider our second model $b^G(z) = b_0^{G}(1+z)$, we measure $b_0^G = 0.98\pm 0.10$. We note that there is a slight dependence on $\ell_{max}$, which could be due to statistical fluctuations, a failure of our linear bias model on small scales, or other effects. However, this dependence is negligible for the purpose of this paper: even in the extreme case of using $\ell_{max} = 2000$ instead of our fiducial 400, the bias we measure is higher by only 12\%, translating into a change in ISW amplitude of $0.13\sigma$.

For the AGN with bias $b^A(z) = b^A_0 [0.53 + 0.289(1+z)^2]$, we measure $b^A_0 = 1.26 \pm 0.23$. Our result is stable with respect to changes in $\ell_{max}$ and is only $\sim 1 \sigma$ higher than the SPT result \cite{Geach13}, $b_0^A = 0.97 \pm 0.13$. Again, this uncertainty corresponds to a shift in the ISW amplitude derived from the AGN sample of about 0.3$\sigma$ and is therefore not important for the purpose of this work.

\section{ISW results}
\label{ISWres}
We measure the cross-correlation of the WISE galaxy and AGN samples with the WMAP CMB temperature maps in the Q, V and W bands. We estimate the signal in 7 bins (bandpowers), equally spaced in $\ell$ space and spanning multipoles from 5 to 100.  Since we are only interested in $\ell \leq 100$, we use maps with HEALPix $N_{side} = 128$, after correcting for the WMAP beam (different for each band) and for the pixel window function. The complex geometry of the mask induces non-trivial off diagonal correlations between bandpowers and we use the MASTER algorithm \cite{Master} to largely undo the effect of the mask and obtain an unbiased (but slightly suboptimal) estimate of the bandpowers.

To estimate the error bars and the covariance matrix, we cross-correlate the WISE galaxy and AGN samples\footnote{Because of the uncertainties on the WISE selection function and noise properties, we choose to use the real data in estimating the Monte Carlo covariance matrix.} with 5000 simulated CMB maps as follows:
We use our fiducial cosmology CMB power spectrum and the WMAP beam transfer function to obtain 5000 simulated CMB maps (Gaussian random fields) for each band. Then noise is added to each pixel in the form of a Gaussian random variable with zero mean and standard deviation given by $\sigma = \sigma_0 / \sqrt{N_{exp}}$, where $\sigma_0$ is 2.188, 3.131 or 6.544 mK, for Q, V and W bands respectively, and $N_{exp}$ is the number of exposures of the corresponding pixel in the WMAP survey.

The Monte Carlo covariance matrices for the Q band are shown in figure \ref{cov} in appendix \ref{append}. We verify convergence by varying the number of simulations and noting consistent results. While the covariance matrix is dominated by the diagonal components, the off diagonal components are non-negligible and should be taken into account.

The cross-correlation results are shown in figures \ref{cross_gals} and \ref{cross_qso} and summarized in tables \ref{tab:gal} and \ref{tab:qso}. If $\d$ are the measured bandpowers and $\t$ are the corresponding theory values, the best fit amplitude $\mathcal{A} = C_{\ell}^{Tg, best fit} / C_{\ell}^{Tg, \Lambda CDM}$ is obtained by minimizing $\chi^2 = (\d - \t)^T C^{-1} (\d - \t)$, where $C^{-1}$ is the inverse of the covariance matrix. The significance is computed as $\sqrt{\chi^2_{null} - \chi^2_{min}}$, with $\chi^2_{null}$ referring to the null hypothesis $\t = \bf{0}$ (i.e. no ISW signal).

The null tests are performed by cross-correlation with the simulated CMB maps are shown in figure \ref{null} in appendix \ref{append} for each band. All of the null tests are consistent with zero signal as expected.

\subsection{Galaxies}

\begin{figure}[ht]
\includegraphics[width=8.5cm]{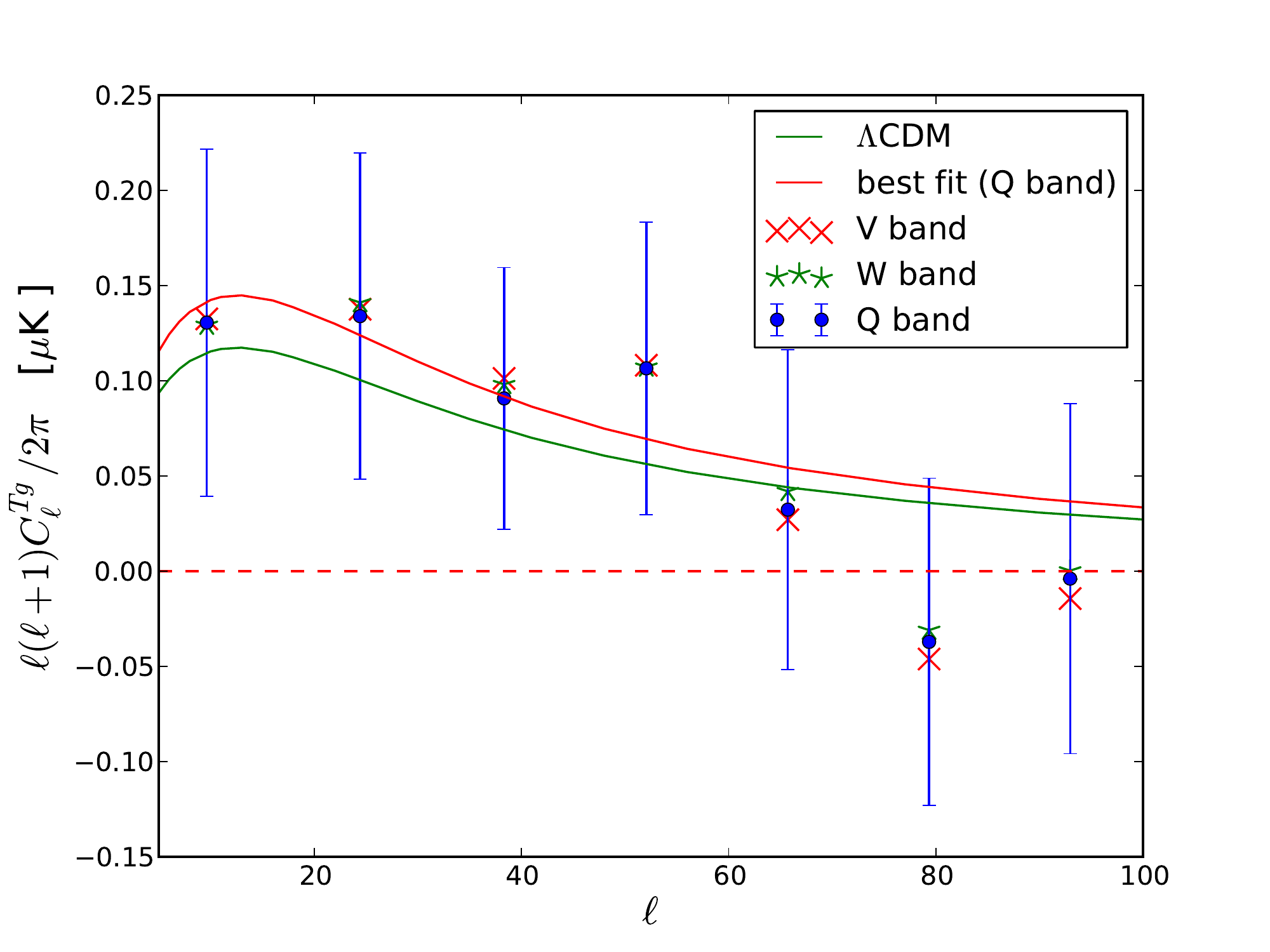}
 \caption{cross-correlation between WISE galaxies and WMAP temperature maps, where the $\Lambda$CDM theory curve is computed assuming a redshift independent bias. The error bands are shown only for Q band and the other error bars are within 5\% of the ones shown.}\label{cross_gals}
\end{figure}

For WISE galaxies with constant bias, we measure an amplitude of $\mathcal{A}  = 1.24\pm 0.47$, fully consistent with the $\Lambda$CDM prediction $\mathcal{A} = 1$.  The amplitudes and some basic statistical properties for each band are reported in table \ref{tab:gal} and the results are shown in figure \ref{cross_gals}.  

The correlation signal is essentially independent of frequency over the range 40 - 90 GHz, which makes a significantly contamination by foregrounds unlikely. Moreover, the amplitude is stable under small changes in the mask, as expected.

\begin{table}[ht]
\begin{tabular}{ | c | c | c | c | c | }
\hline
  Band & Amplitude $\mathcal{A}$ & $\chi^2$ / dof & p-value & \ $S/N$ \ \\
\hline \hline
  Q & 1.22 $\pm$ 0.47 & 1.42 / 6 & 0.04 & 2.6  \\
  \hline
  V & 1.25 $\pm$ 0.47 & 1.81 / 6 & 0.06 & 2.6  \\
  \hline
  W & 1.26 $\pm$ 0.47 & 1.25 / 6 & 0.03 & 2.7 \\
  \hline
\end{tabular}
\caption{ISW amplitude and significance for the galaxy sample, assuming a constant bias model.}
\label{tab:gal}
\end{table}

As we can see from table \ref{tab:gal}, the $\chi^2$ of the best fit slightly low, but it is expected this high or low about 6 - 12\% of the time. 
To test the error calculation, we used the Gaussian approximation (Fisher formalism) to analytically compute the errors bars using the measured WISE auto-power spectrum, obtaining a result that is fully consistent with the Monte Carlo estimate.

To assess the dependence of our result on uncertainties in the evolution of bias, we repeat the analysis with a model in which it evolves linearly with redshift $b^G(z) = b_0^{G}(1+z)$. In this case we find $\mathcal{A} = 1.54 \pm 0.59$, with again $S/N \approx 2.7$. This corresponds to a shift in amplitude of about one half sigma and therefore we can conclude that our measurement is fairly robust under uncertainties in the evolution of the bias.

\subsection{AGN}
\begin{figure}[ht]
\includegraphics[width=8.5cm]{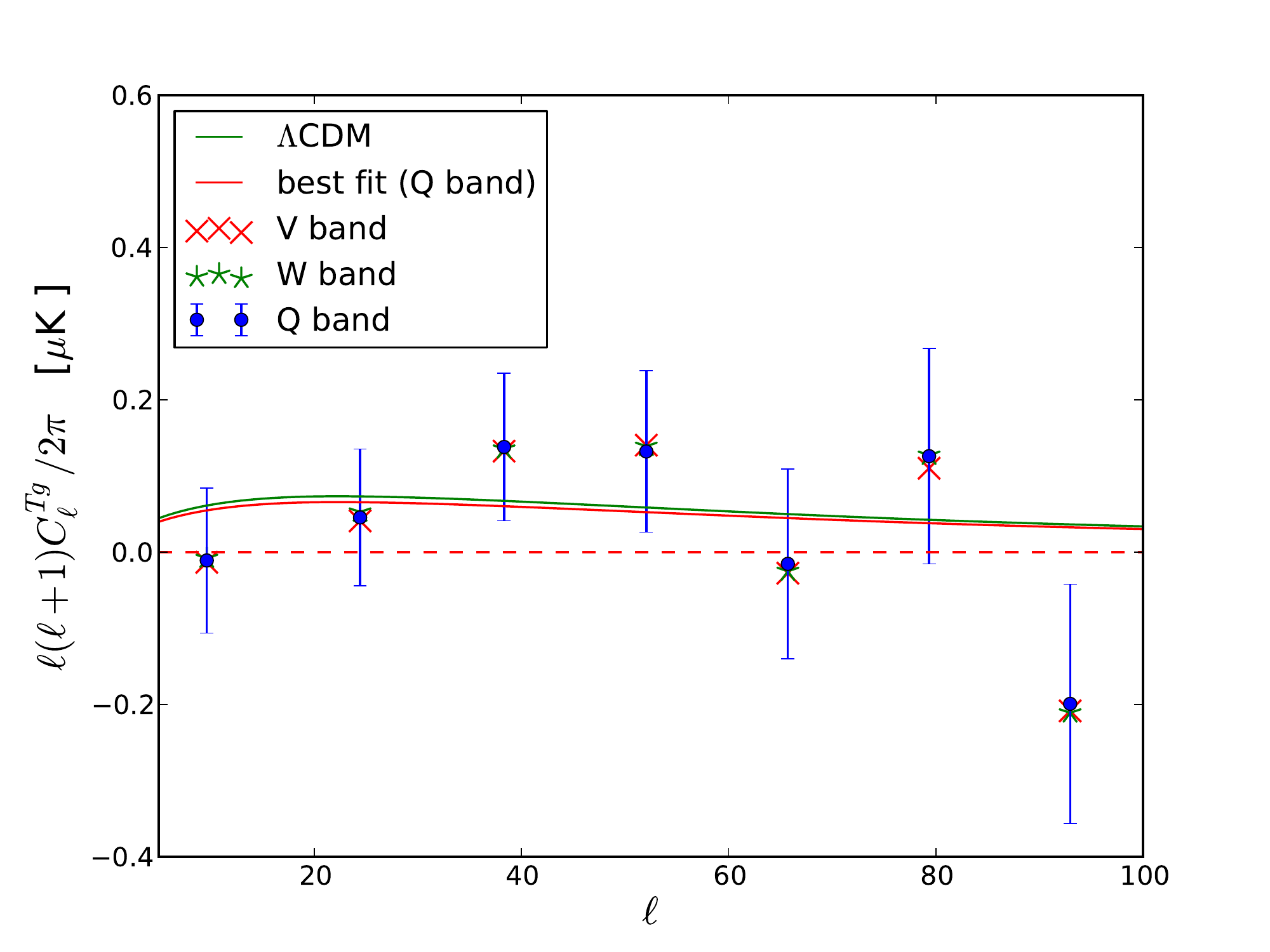}
 \caption{Cross-correlation between WISE AGN and WMAP temperature maps. The error bands are shown only for Q band; the other error bars are within 5\% of the ones shown.}\label{cross_qso}
\end{figure}

The measured amplitude $\mathcal{A} = 0.88 \pm 0.74$ is again consistent with the $\Lambda$CDM predictions. The amplitude is stable under small changes in the mask and is frequency independent, as can be seen from table \ref{tab:qso} and figure \ref{cross_qso}.

\begin{table}[ht]
\begin{tabular}{ | c | c | c | c | c | }
\hline
  Band & Amplitude $\mathcal{A}$ & $\chi^2$ / dof & p-value & \ $S/N$ \ \\
\hline \hline
  Q & 0.88 $\pm$ 0.74 & 4.6 / 6 & 0.4 & 1.2  \\
  \hline
  V & 0.86 $\pm$ 0.74 & 4.7 / 6 & 0.4 & 1.2  \\
  \hline
  W & 0.91 $\pm$ 0.75 & 4.6 / 6 & 0.4 & 1.2  \\
  \hline
\end{tabular}
\caption{ISW amplitude and significance for the AGN sample.}
\label{tab:qso}
\end{table}

As the highest $\ell$ bin in figure \ref{cross_qso} appears low, we extend our analysis to $\ell = 200$ to test that this value is simply a fluctuation. We find that the points for $100 \leq \ell \leq 200$ are consistent with the low-$\ell$-fit signal curve as expected.

Though the AGN sample is simpler to cleanly select than the galaxy sample, the significance of the AGN sample is lower. Partially this is because the expected signal itself is smaller, as a large fraction of WISE AGN lie at $z \agt 1$, where Dark Energy is unimportant. In addition, the smaller number of sources makes the sample shot-noise limited in the high $\ell$ bins, further reducing the significance of the measurement.

\section{Conclusions}
\label{sec:conclusions}
Dark Energy remains one of the most elusive outstanding problems in Physics, and the ISW effect provides one of the most direct probes of its properties.

In this work we have measured the cross-correlation between the CMB temperature and both WISE galaxies and AGN.  The correlation is expected to be entirely due to the ISW effect and hence absent in a Universe with no Dark Energy.

We find a positive signal which is consistent with the $\Lambda$CDM predictions, with significances of 2.7$\sigma$ and 1.2$\sigma$ for galaxies and AGN respectively and a combined significance close to $3\sigma$. It can be shown that the bulk contribution to the galaxy ISW signal comes from $z \sim 0.2 - 0.6$, with a peak at $z \sim 0.3$, while the AGN, due to their fairly high median redshift, receive a fairly uniform contribution in the interval $z \sim 0.2 - 1.2$. Therefore, the AGN act as a useful probe of Dark Energy at an earlier time. We find that our results show no evidence for evolution of the Dark Energy density, as expected from a Cosmological Constant.

We use CMB lensing potential from the Planck mission to calibrate the bias and stellar contamination of our sample, a method that has recently become available with advances in high-resolution CMB experiments. Calibration with lensing cross-correlation allows a direct measurement of the effective bias, with a smaller sensitivity to some systematic errors that can affect a measurement with the auto power spectra.

The signal we detect is independent of the choice of mask and, more crucially, frequency independent. An imperfect foreground subtraction on the CMB side could potentially create spurious correlation with WISE, but any residual foreground contamination is expected to vary significantly over the range 40 - 90 GHz that we probe here. Therefore we conclude that any contamination, if present, is likely to be highly subdominant.

While some previous studies hinted at the possibility of a signal with amplitude higher than what expected from $\Lambda$CDM, we find no deviation from the standard cosmological model in either amplitude or redshift dependence, in agreement with some of the other previous measurements (eg \cite{Scranton, Gian12, Gian13, PlanckISW}). We are also in agreement with a recent analysis of the ISW effect from WISE galaxies \cite{Goto2} that used a somewhat smaller sample at lower redshift and measured an amplitude consistent with $\Lambda$CDM, with a significance of about $1\sigma$. 
\ 
\\

{\em Acknowledgements.}
We thank Michael Strauss, Kendrick Smith, Fabian Schmidt, Olivier Dor\'e, Eiichiro Komatsu, Amir Hajian and Matias Zaldarriaga for very helpful discussions.
SF and DNS are supported by NASA ATP grant NNX12AG72G. BDS was supported by a Miller Research Fellowship at Berkeley and by a Charlotte Elizabeth Procter Honorific Fellowship at Princeton University.
\appendix
\label{append}
\section{Covariance matrix and null tests}
Here we show plots for the Q-band covariance matrices and the null tests. For a description of the methodology, see section \ref{ISWres} in the main text.

\begin{figure}[ht]
 \subfigure{\includegraphics[width=8.5cm]{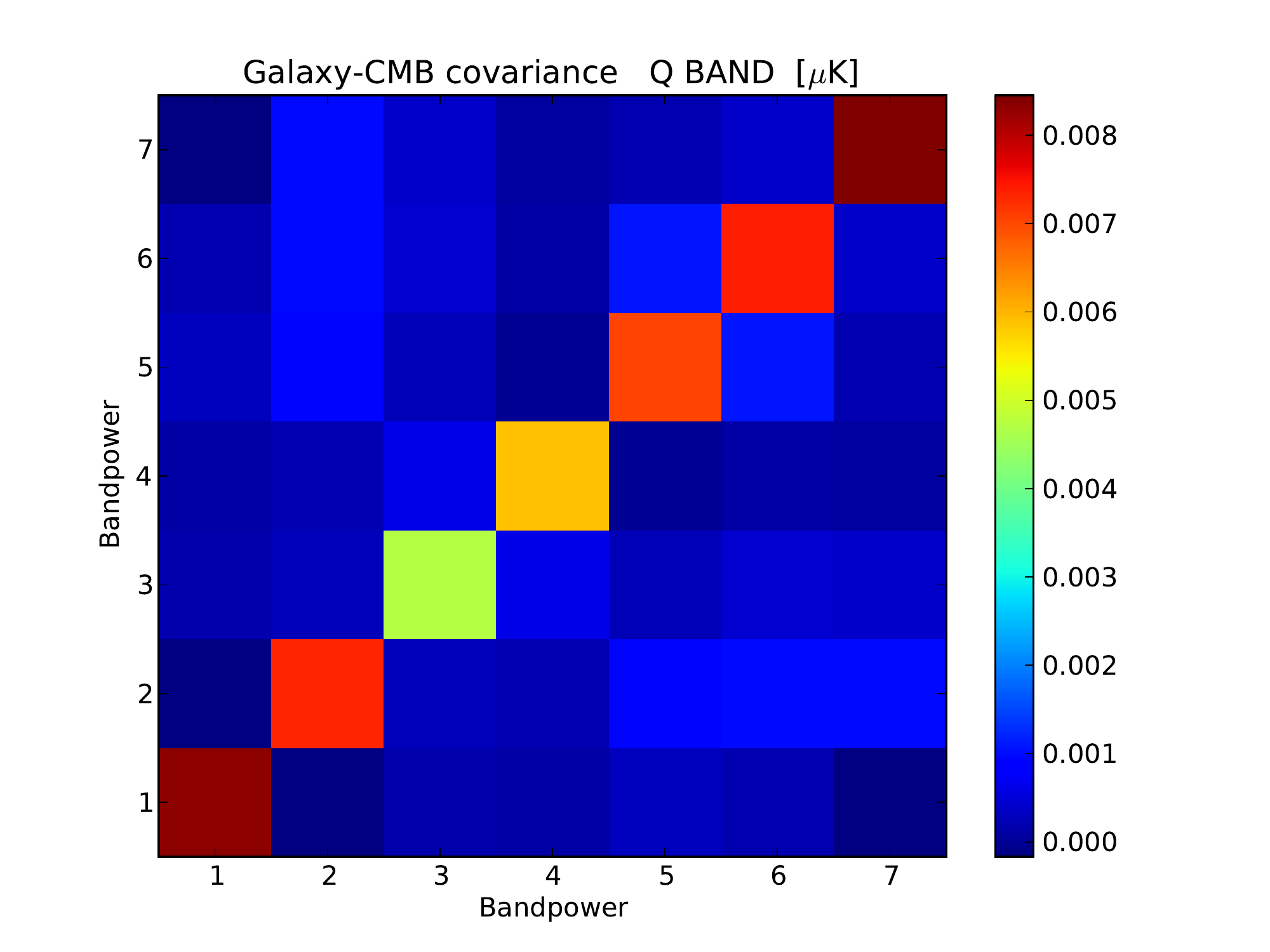}} 
 \subfigure{\includegraphics[width=8.5cm]{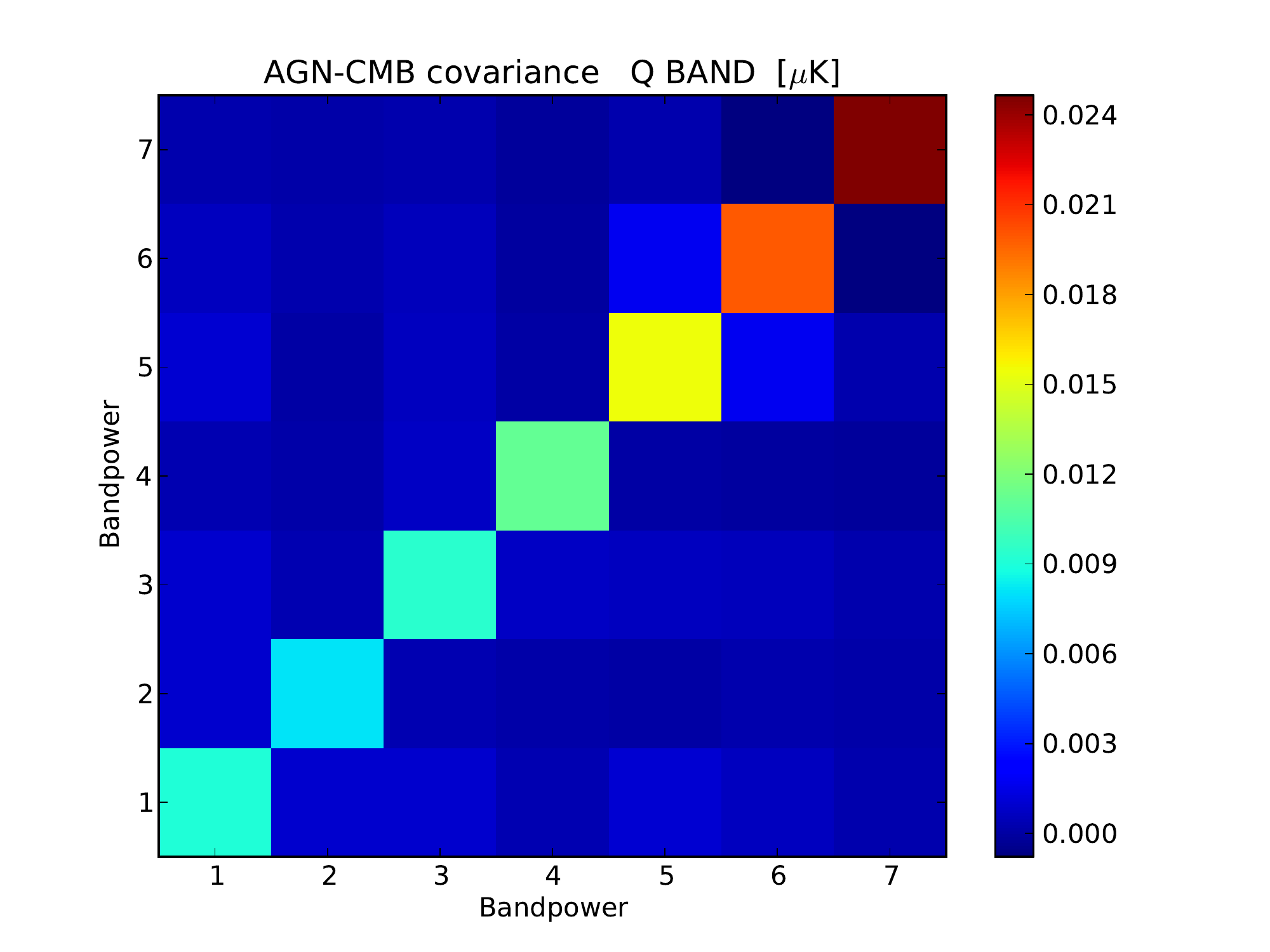}}
 \caption{Q-band Monte Carlo covariance matrices for galaxies (top) and AGN (bottom). V and W band covariances are very similar and are not shown here.}\label{cov} 
\end{figure}\begin{figure}[ht]
 \subfigure{\includegraphics[width=8.5cm]{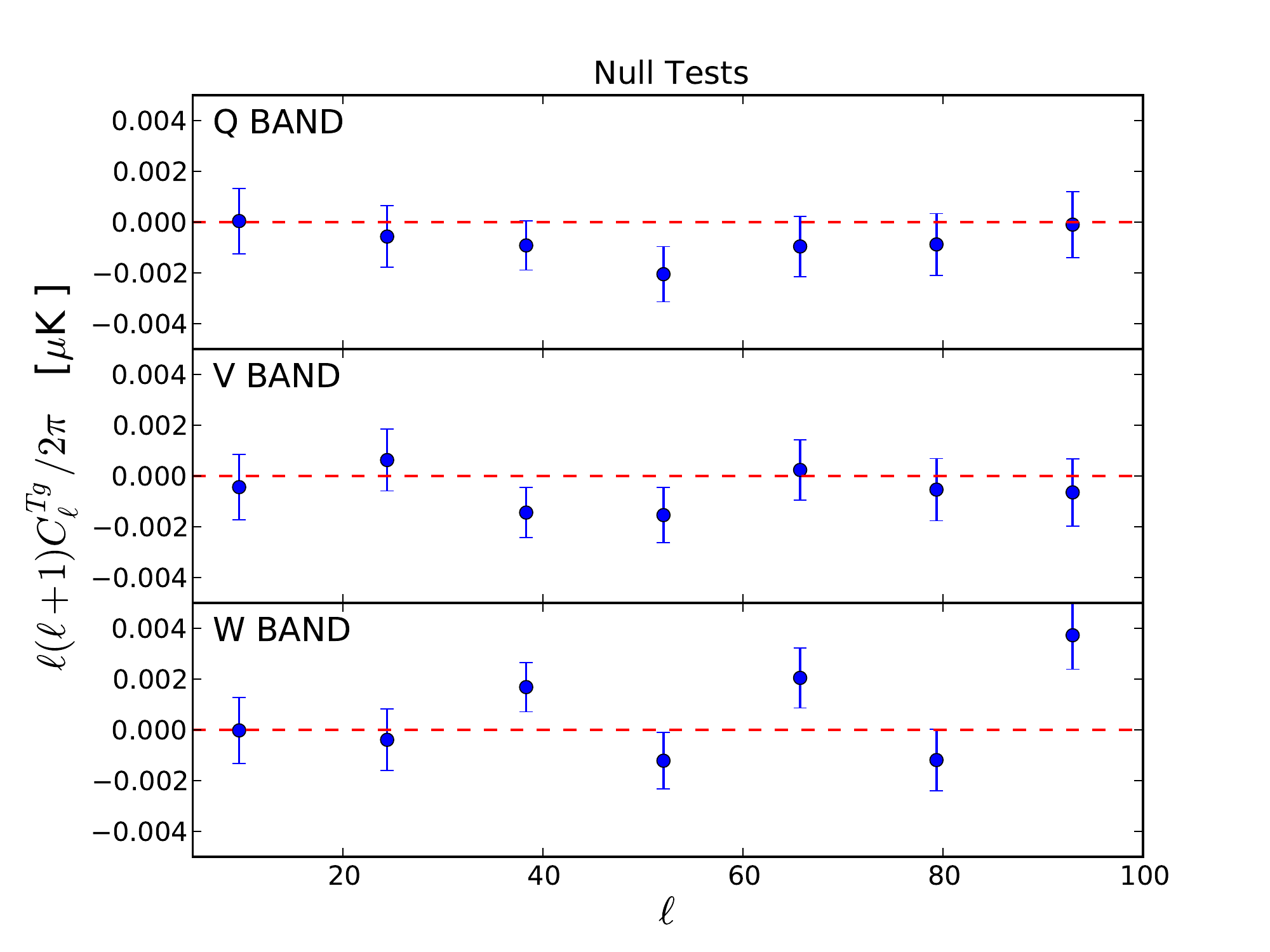}} 
 \subfigure{\includegraphics[width=8.5cm]{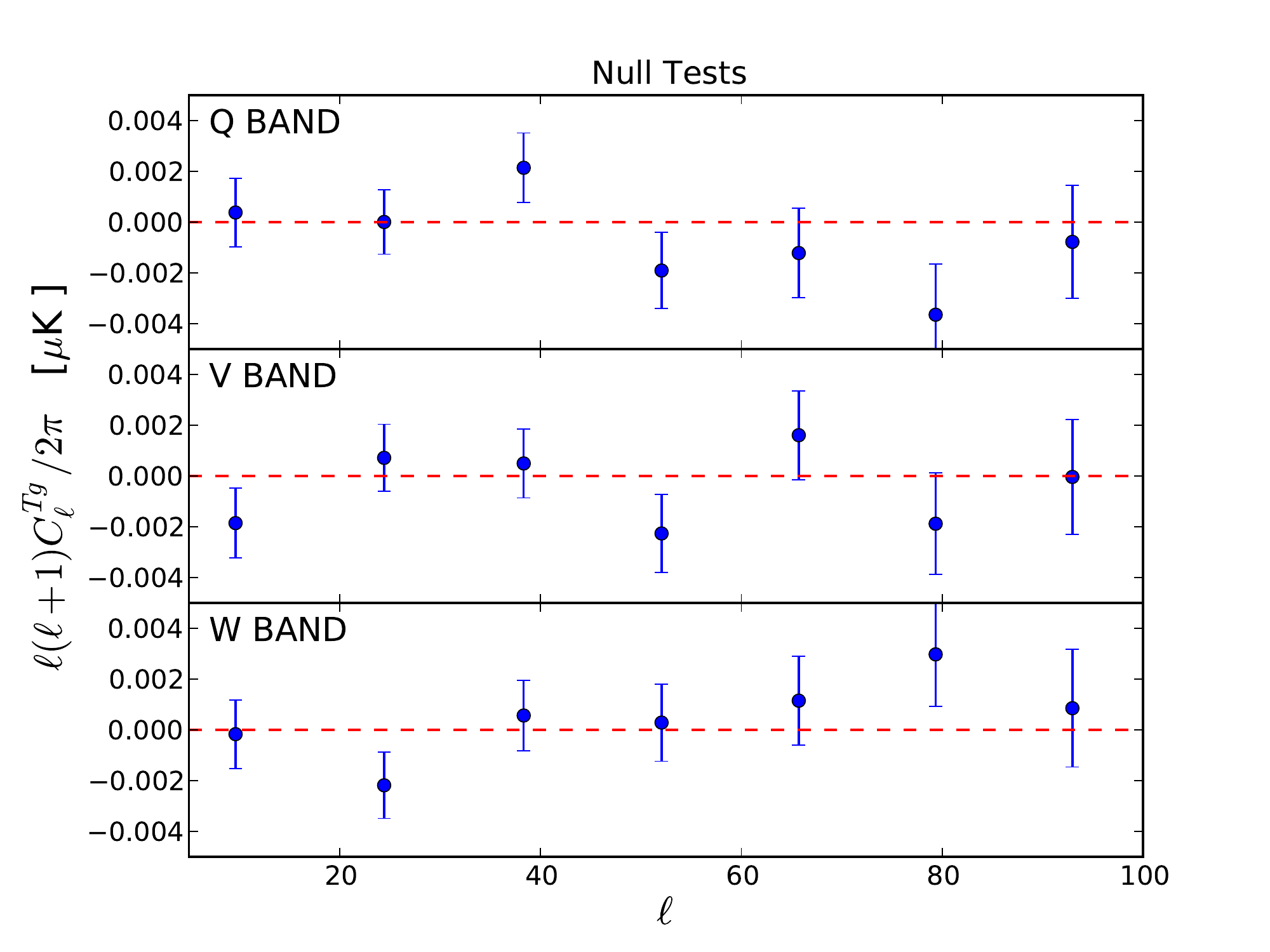}}
 \caption{Null tests: correlation of the WISE galaxy (top) and AGN (bottom) maps with 5000 simulated CMB realizations. All are consistent with zero signal.}\label{null} 
\end{figure}

\newpage
\bibliographystyle{prsty}

\end{document}